\documentclass[prb,twocolumn,showpacs,preprintnumbers,amsmath,amssymb]{revtex4}

\usepackage{graphicx}
\usepackage{dcolumn}
\usepackage{bm}

\begin{document}


\title{Novel superconductivity on the magnetic criticality in heavy-fermion systems : a systematic study of NQR under pressure}

\author{Y.~Kitaoka}%

\author{S.~Kawasaki}%

\author{Y.~Kawasaki}%
\altaffiliation[Present address : ]{Department of Physics, Faculty of Engineering, Tokushima University, Tokushima 770-8506, Japan}%

\author{T.~Mito}%
 \altaffiliation[Present address : ]{Department of Physics, Faculty of Science, Kobe University, Hyogo 657-8501, Japan}%

\author{G.-q.~Zheng}%
 \altaffiliation[Present address : ]{Department of Physics, Faculty of Science, Okayama University, Okayama 700-8530, Japan}%

\affiliation{Department of Materials Science and Technology, Graduate School of Engineering Science, Osaka University,  Osaka 560-8531, Japan}


\date{\today}

\begin{abstract}
We report the discovery of exotic superconductivity (SC) and novel magnetism in heavy-fermion (HF) compounds, CeCu$_2$Si$_2$, CeRhIn$_5$ and CeIn$_3$ on the verge of antiferromagnetism (AFM) through nuclear-quadrupole-resonance (NQR) measurements under pressure ($P$). The exotic SC in a homogeneous CeCu$_2$Si$_2$ ($T_{\rm c}=0.7$ K) revealed {\it antiferromagnetic critical fluctuations} at the border to AFM or marginal AFM. Remarkably, it has been found that the application of magnetic field induces an spin-density-wave (SDW) transition by suppressing the SC near the upper critical field.
Furthermore, the uniform mixed phase of SC and AFM in CeCu$_2$(Si$_{1-x}$Ge$_x$)$_2$ emerges on a microscopic level, once a tiny amount of 1\%Ge($x=0.01$) is substituted for Si to expand its lattice. The application of minute pressure ($P\sim 0.19$ GPa) suppresses the sudden emergence of the AFM caused by doping Ge. The persistence of the low-lying magnetic excitations at temperatures lower than $T_c$ and $T_N$ is ascribed due to the uniform mixed phase of SC and AFM.

Likewise, the $P$-induced HF superconductor CeRhIn$_5$ coexists with AFM on a microscopic level in $P = 1.5$ - 1.9 GPa. It is demonstrated that SC does not yield any trace of gap opening in low-lying excitations below the onset temperature, presumably associated with {\it an amplitude fluctuation of superconducting order parameter}. The unconventional gapless nature of SC in the low-lying excitation spectrum emerges due to the uniform mixed phase of AFM and SC.

By contrast, in CeIn$_3$, the $P$-induced {\it phase separation} into AFM and paramagnetism (PM) takes place without any trace for a quantum phase transition. The outstanding finding is that SC sets in at both the phases  magnetically separated into AFM and PM in $P=2.28-2.5$ GPa. A new type of SC forms the uniform mixed phase with the AFM and the HF SC takes place in the PM.  We propose that the magnetic excitations such as spin-density fluctuations induced by the first-order phase transition from the AFM to the PM  might mediate attractive interaction to form the Cooper pairs in the novel phase of AFM. 
\end{abstract}

\maketitle

\section{Introduction}

The most common kind of superconductivity (SC) is based on bound electron pairs coupled by deformation of the lattice. However, SC of more subtle origins is rife in strongly correlated electron systems including many heavy-fermion (HF), cuprate and organic superconductors. In particular, a number of studies on $f$-electron compounds revealed that unconventional SC arises at or close to a quantum critical point (QCP), where magnetic order disappears at low temperature ($T$) as a function of lattice density via application of hydrostatic pressure ($P$) \cite{Jaccard92,Movshovic96,Mathur98,Hegger00}. These findings suggest that the mechanism forming Cooper pairs can be magnetic in origin. Namely, on the verge of magnetic order, the magnetically soft electron liquid can mediate spin-dependent attractive interactions between the charge carriers \cite{Mathur98}.  
However, the nature of SC and magnetism is still unclear when SC appears very close to the antiferromagnetism (AFM). Therefore, in light of an exotic interplay between these phases, unconventional electronic and magnetic properties around QCP have attracted much attention and a lot of experimental and theoretical works are being extensively made.

\begin{figure}[htbp]
\centering
\includegraphics[width=7.1cm]{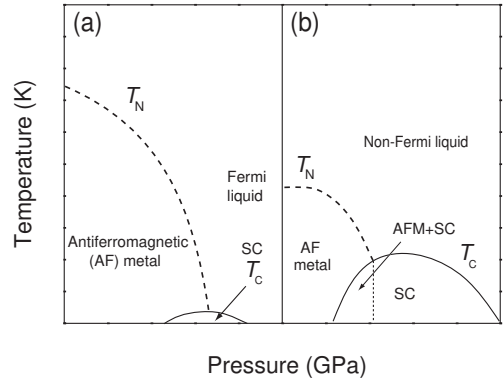}
\caption[]{Schematic phase diagrams of HF compounds: (a) CePd$_2$Si$_2$ \cite{Mathur98,Grosche96,Grosche01}, CeIn$_3$ \cite{Mathur98,Grosche01,Walker97,Muramatsu01,Knebel02} and CeRh$_2$Si$_2$ \cite{Movshovic96,Araki}: (b) CeCu$_2$Si$_2$ \cite{Steglich,Bellarbi,Bellarbi2,kawasaki01,Ykawasaki2} and CeRhIn$_5$ \cite{Hegger00,Muramatsu}. Dotted and solid lines indicate the $P$ dependence of $T_N$ and $T_c$, respectively.}
\end{figure}

The phase diagram, schematically shown in figure 1(a), has been observed in antiferromagnetic HF compounds such as CePd$_2$Si$_2$ \cite{Mathur98,Grosche96,Grosche01}, CeIn$_3$ \cite{Mathur98,Grosche01,Walker97,Muramatsu01,Knebel02}, and CeRh$_2$Si$_2$ \cite{Movshovic96,Araki}. Markedly different behavior, schematically shown in figure 1(b), has been found in the archetypal HF superconductor CeCu$_2$Si$_2$ \cite{Steglich,Bellarbi,Bellarbi2,kawasaki01,Ykawasaki2} and the more recently discovered CeRhIn$_5$ \cite{Hegger00,Muramatsu}. Although an analogous behavior relevant to a magnetic QCP has been demonstrated in these compounds, it is noteworthy that the associated superconducting region extends to higher densities than in the other compounds; their value of $T_c$ reaches its maximum away from the verge of AFM \cite{Bellarbi,Bellarbi2,Muramatsu}.

In this article, we review the recent studies under $P$ on CeCu$_2$Si$_2$, CeRhIn$_5$ and CeIn$_3$ via nuclear-quadrupole-resonance (NQR) measurements. These systematic works have revealed the homogeneous mixed phase of SC and AFM and that its novel superconducting nature exhibits the gapless nature in the low-lying excitations below $T_c$, which differ from the superconducting characteristics for the HF superconductors reported to possess the line-node gap \cite{Ykawasaki2,Mito01,Shinji03}. 


\section{Exotic magnetism and superconductivity on the magnetic criticality in CeCu$_2$Si$_2$}

\subsection{The temperature versus pressure phase diagram}

The firstly-discovered HF superconductor CeCu$_2$Si$_2$ is located just at the border to the AFM at $P=0$ \cite{ishida99}.
This was evidenced by various magnetic anomalies observed above $T_c$ \cite{nakamura,Steglich} and by the fact that the magnetic {\it A-phase} appears when SC is suppressed by a magnetic field $H$ \cite{Bruls94}.
Furthermore, the transport, thermodynamic and NQR measurements  consistently indicated that nominally off-tuned Ce$_{0.99}$Cu$_{2.02}$Si$_2$ is located just at $P_c$ and crosses its QCP by applying a minute pressure of $P=0.2$ GPa \cite{gegenwart98,kawasaki01}. The magnetic and superconducting properties in CeCu$_2$Si$_2$ were investigated around the QCP as the functions of $P$ for Ce$_{0.99}$Cu$_{2.02}$Si$_2$ just at the border to AFM and of Ge content $x$ for CeCu$_2$(Si$_{1-x}$Ge$_x$)$_2$ by Cu-NQR measurements \cite{kawasaki01,kawasaki02}. Figure 2 shows the phase diagram referred from the literature \cite{kawasaki02}.
Here, $T_F^*$ is an effective Fermi temperature below which the nuclear-spin-lattice-relaxation rate divided by temperature ($1/T_1T$) stays constant and $T_m$ is a temperature below which the slowly fluctuating antiferromagnetic waves start to develop.
Note that a primary effect of Ge doping expands the lattice \cite{trovarelli97} and that its chemical pressure is $-0.076$ GPa per 1\% Ge doping as suggested from the $P$ variation of Cu-NQR frequency $\nu_Q$ in CeCu$_2$Ge$_2$ and CeCu$_2$Si$_2$ \cite{kitaoka95}.

\begin{figure}[tbp]
\centering
\includegraphics[width=8.3cm]{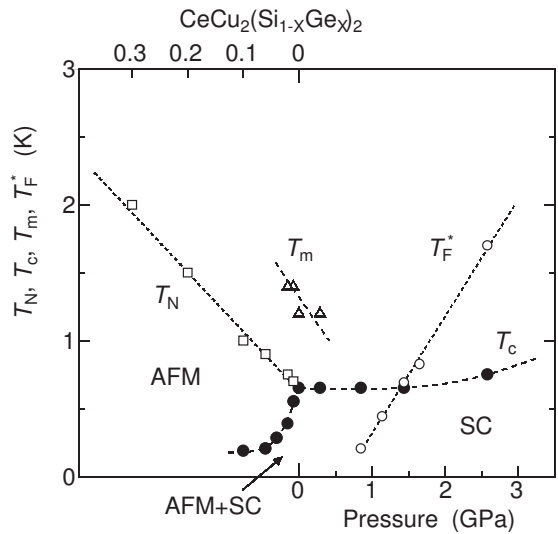}
\caption[]{The combined phase diagram of AFM and SC for CeCu$_2$(Si$_{1-x}$Ge$_{x}$)$_2$ and for Ce$_{0.99}$Cu$_{2.02}$Si$_2$ under $P$.
$T_N$ and $T_c$ are the respective transition temperature of AFM and SC\@.
Also shown are $T_m$ below which the slowly fluctuating AFM waves develop and $T_F^*$ below which $1/T_1T$ becomes const.
}
\label{fig:2}
\end{figure}

In the normal state, the slowly fluctuating antiferromagnetic waves propagate over a long-range distance without any trace of AFM below $T_m \sim 1.2$ K\@. The exotic SC emerges in Ce$_{0.99}$Cu$_{2.02}$Si$_2$ below $T_c \sim 0.65$ K, where low-lying magnetic excitations remain active even below $T_c$. A rapid decrease below $T_c$ in $1/T_1$ evidences the opening of superconducting energy gap, whereas the large enhancement in $1/T_1T$ well below $T_c$ reveals the gapless nature in the low-lying excitations in its superconducting state \@. With increasing $P$, as a result of the marked suppression of antiferromagnetic critical fluctuations, the exotic SC evolves into a typical HF-SC with the line-node gap that is characterized by the relation of $1/T_1\propto T^3$ above $P=0.85$ GPa\@.

\subsection{The uniform mixed phase of AFM and SC in CeCu$_2$(Si$_{1-x}$Ge$_x$)$_2$}

Markedly by substituting only 1\% Ge, AFM emerges at $T_N \sim$ 0.7 K, followed by the SC at $T_c \sim$ 0.5 K \@. Unexpectedly, $1/T_1$ does not show any significant reduction at $T_c$, but follows a $1/T_1T$ = const.\ behavior well below $T_c$ as observed in Ce$_{0.99}$Cu$_{2.02}$Si$_2$\@ as presented in Fig.3.  It was revealed that the uniform mixed phase of SC and AFM is unconventional, exhibiting that low-lying magnetic excitations remain active even below $T_c$ \@ as shown later on Fig.4. As Ge content increases, $T_N$ is progressively increased, while $T_c$ is steeply decreased. As a result of the suppression of  antiferromagnetic critical fluctuations for the samples at more than $x=0.06$, the magnetic properties above $T_N$ progressively change to those in a localized regime as observed in CeCu$_2$Ge$_2$\@ \cite{kitaoka95}.

Further insight into the exotic SC is obtained on CeCu$_2$(Si$_{0.98}$Ge$_{0.02}$)$_2$ that reveals the uniform mixed phase of AFM ($T_N \sim$ 0.75 K) and SC ($T_c \sim$ 0.4 K) under $P = 0$. 
 Figure~3 shows the $T$ dependence of $1/T_1$ at $P$ = 0 GPa (closed circles), 0.56 GPa (open circles) and 0.91 GPa (closed squares). Here, the data at $P$ = 0.19 GPa are not shown, since they are nearly equivalent to those at $P$ = 0.56 GPa\@.
In the entire $T$ range, $1/T_1$ is suppressed with increasing $P$, evidencing that the low-energy component of spin fluctuations is forced to shift to a high-energy range.
As expected from the fact that the AFM is already suppressed at pressures exceeding $P = 0.19$ GPa, any trace of anomaly associated with it is not observed at all down to $T_c \sim$ 0.45 K at $P$ = 0.56 GPa and 0.91 GPa\@.
It is, therefore, considered that the AFM in CeCu$_2$(Si$_{0.98}$Ge$_{0.02}$)$_2$ is not triggered by some disorder effect but by the intrinsic lattice expansion due to the Ge doping.

\begin{figure}[htbp]
\centering
\includegraphics[width=8cm]{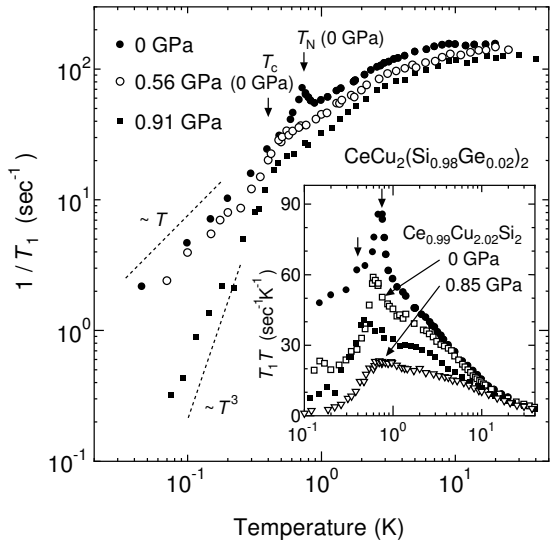}
\caption[]{The $T$ dependence of $1/T_1$ of CeCu$_2$(Si$_{0.98}$Ge$_{0.02}$)$_2$ at several pressures.
Inset shows the $T$ dependence of $1/T_1T$ of CeCu$_2$(Si$_{0.98}$Ge$_{0.02}$)$_2$ at $P$ = 0 GPa (closed circles) and 0.91 GPa (closed squares) and those of Ce$_{0.99}$Cu$_{2.02}$Si$_2$ at $P$ = 0 (open squares) and 0.85 GPa (open triangles).
Arrows indicate $T_N \sim 0.75$ K and $T_c \sim$ 0.4 K at $P = 0$ GPa for CeCu$_2$(Si$_{0.98}$Ge$_{0.02}$)$_2$\@.
}
\label{fig:1}
\end{figure}

In order to demonstrate a systematic evolution of low-energy magnetic excitations at the paramagnetic state, the inset of figure 3 shows the $T$ dependence of $1/T_1T$ in CeCu$_2$(Si$_{0.98}$Ge$_{0.02}$)$_2$ at $P$ = 0 GPa (closed circles) and 0.91 GPa (closed squares) along with the results in Ce$_{0.99}$Cu$_{2.02}$Si$_2$ at $P$ = 0 GPa (open squares) and 0.85 GPa (open triangles) \cite{kawasaki01,kawasaki02}.
The  result of $1/T_1T$ in CeCu$_2$(Si$_{0.98}$Ge$_{0.02}$)$_2$ at $P = 0$ GPa is well explained by the spin-fluctuations theory for weakly itinerant AFM   in $T_c < T < 1.5$ K around $T_N \sim$ 0.75 K \cite{kitaoka01,kawasaki01,kawasaki02}.
The good agreement between the experiment and the calculation indicates that a long-range nature of the AFM is in the itinerant regime.
At $P = 0.91$ GPa, $1/T_1T$, which probes the development of magnetic excitations, is suppressed and resembles a behavior that would be expected at an intermediate pressure between $P = 0$ GPa and 0.85 GPa for Ce$_{0.99}$Cu$_{2.02}$Si$_2$\@.

Next, we discuss an intimate $P$-induced evolution of low-lying magnetic excitations in the superconducting state.
As seen in figure 3 and its inset, the $1/T_1$ and $1/T_1T$ at $P =$ 0 GPa do not show a distinct reduction below $T_c$, but instead, a $T_1T=$ const.\ behavior emerges well below $T_c$\@.
At $P$ = 0.56 GPa, the AFM is depressed, but anftiferromagnetic critical fluctuations develop in the normal state. It is noteworthy that the relation of $1/T_1\propto T$ is still valid below $T_c$, resembling the behavior for Ce$_{0.99}$Cu$_{2.02}$Si$_2$ at $P = 0$ GPa \cite{kawasaki01,ishida99}.
By contrast at $P = 0.91$ GPa, $1/T_1$ follows a relation of $1/T_1 \propto T^3$ below $T_c \sim$ 0.45 K, consistent with the line-node gap at the Fermi surface.
This typical HF-SC in $1/T_1$ was observed in Ce$_{0.99}$Cu$_{2.02}$Si$_2$ at pressures exceeding $P=$ 0.85 GPa as well \cite{kawasaki01}.
A small deviation from $1/T_1\propto T^3$ behavior at $P = 0.91$ GPa far below $T_c$ may be associated with an inevitable Ge-impurity effect for $d$-wave superconductors in general \cite{schmitt-rink86,hotta93}.
Therefore, it is considered that the unconventional SC at $P = 0$ GPa and 0.56 GPa evolves into the typical HF-SC with the line-node gap at pressures exceeding $P = $0.91 GPa\@.
Apparently, these results exclude a possible impurity effect as a primary cause for the $T_1T$ = const.\ behavior below $T_c$ at $P =$ 0 GPa\@.
We stress that a reason why the $1/T_1$ at $P = 0$ GPa is deviated from $1/T_1\propto T^3$ below $T_c$ is ascribed not to the impurity effect but to the persistence of low-lying magnetic excitations well below $T_c$.

\subsection{Evidence for antiferromagnetic critical fluctuations}

Figure 4 indicates the $T$ dependence of NQR intensity multiplied by temperature $I(T)\times T$ in CeCu$_2$(Si$_{0.98}$Ge$_{0.02}$)$_2$ at $P =$ 0, 0.19, 0.56, and 0.91 GPa\@.
Here, the $I(T)$ normalized by the value at 4.2 K is an integrated intensity over frequencies where NQR spectrum was observed.
Note that $I(T)\times T$ stays constant generally, if $T_1$ and/or $T_2$ range in the observable time window that is typically more than several microseconds.
Therefore, the distinct reduction in $I(T)\times T$ upon cooling is ascribed to the development of antiferromagnetic critical fluctuations, since it leads to an extraordinary short relaxation time of $\sim$ 0.14 $\mu$sec \cite{ishida99}.
\begin{figure}[htbp]
\centering
\includegraphics[width=8cm]{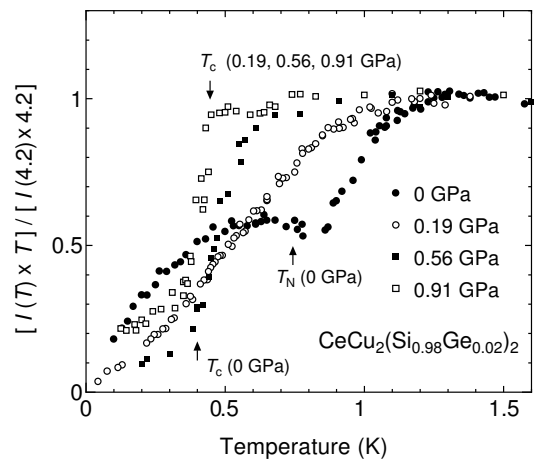}
\caption[]{The $T$ dependence of $I(T)\times T$ at several pressures, where $I(T)$ is an NQR intensity normalized by the value at 4.2 K\@.
}
\label{fig:4}
\end{figure}
The $I(T)\times T=1$ at $P = 0$ GPa decreases down to about $I(T)\times T=0.55$ at $T_N \sim 0.75$ K upon cooling below $T_m \sim$ 1.2 K\@.
Its reduction stops around $T_N$, but does no longer recover with further decreasing $T$\@.
Note that its reduction below $T_c \sim 0.4$ K is due to the superconducting diamagnetic shielding of rf field for the NQR measurement.
As $P$ increases, $T_m$ becomes smaller, in agreement with the result presented in the phase diagram of Fig.~2, and the reduction in $I(T)\times T$  becomes moderate in the normal state.
With further increasing $P$ up to 0.91 GPa, eventually, $I(T)\times T$ remains nearly constant down to $T_c \sim 0.45$ K, indicative of no anomaly related to antiferromagnetic critical fluctuations.
This behavior resembles the result observed at pressures exceeding 
$P = 0.85$ GPa in Ce$_{0.99}$Cu$_{2.02}$Si$_2$\@.
These results also assure that the Ge substitution expands the lattice of Ce$_{0.99}$Cu$_{2.02}$Si$_2$\@.
Thus, the exotic SC in Ce$_{0.99}$Cu$_{2.02}$Si$_2$ and CeCu$_2$(Si$_{0.98}$Ge$_{0.02}$)$_2$ at $P = 0$ GPa is characterized by the persistence of low-lying antiferromagnetic critical excitations.
It was argued that these excitations may be related to a collective mode in the uniform mixed phase of AFM and SC \cite{kitaoka01,kawasaki01,kawasaki02}.
It seems, therefore, that these exotic SC could be rather robust against the appearance of AFM.
\begin{figure}[htbp]
\centering
\includegraphics[width=6.7cm]{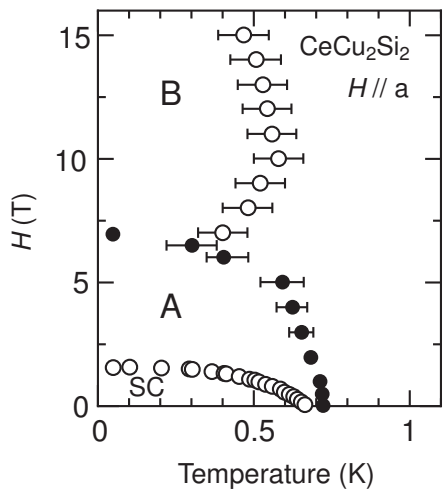}
\caption[]{The magnetic field ($H$) vs $T$ phase
diagram of CeCu$_2$Si$_2$.\cite{Bruls94}}
\label{fig:3}
\end{figure}
\subsection{The magnetic field versus temperature phase diagram}

Here, we deal with the $H$ vs $T$ phase diagram of Ce$_{0.99}$Cu$_2$Si$_2$ shown in figure 5.
When $H$ suppresses its SC, exceeding an upper critical field $H_{c2}$, various measurements revealed an evolution from the SC into some magnetic phase that was called as {\it A-phase} \cite{nakamura,Bruls94,gegenwart98}. This phase emerges below $T_A$ close to the $T_c$ at $H=0$. Recent neutron diffraction experiment has revealed that a single crystal exhibiting {\it A-phase} anomalies below $T_A\sim 0.8$ K undergoes a long-range incommensurate AFM at $H=0$. It was suggested that a spin-density-wave (SDW) instability is the origin of the magnetic QCP in CeCu$_2$Si$_2$ \cite{Stockert}.  Markedly, the recent Cu-NMR measurement is consistent with the onset of static magnetic order as the nature of $H$ induced {\it A-phase} \cite{Nomura}, although {\it B-phase} is not yet under detailed investigations. Upon accepting the existence of a second order quantum critical point $T = 0$ between the uniform mixed phase of SC+SDW and the phase of SC, a promising theoretical approach is available for the analysis of experiments in the study of the influence of an applied magnetic field on such the critical point \cite{Sachdev}. In order to address an origin of {\it A-phase}, further extensive works are required and now in progress.

\subsection{Towards a new concept for superconductivity}

Antiferromagnetic critical fluctuations develop below $T_{\rm m}$ in $0 \leq x < 0.06$ in CeCu$_2$(Si$_{1-x}$Ge$_x$)$_2$ and $0 < P < 0.2$ GPa in CeCu$_2$Si$_2$ \@. Remarkably this marginal AFM emerges closely to the border between AFM and SC. Once slight Ge is substituted for Si to expand its lattice, the AFM suddenly sets in\@.
By contrast, AFM is not observed down to 0.012 K at $x = 0$.
With increasing $x$, $T_{\rm N}$ is progressively increased, while $T_{\rm c}$ is steeply decreased. Correspondingly, antiferromagnetic critical fluctuations are suppressed for the samples at more than $x$ = 0.06.
It is noteworthy that the AFM seems to suddenly disappear at $x$ = 0 as if $T_{\rm N}$ was replaced by $T_{\rm c}$\@.
Eventually, the SC coexists with the marginal AFM  
at $x = 0$ \cite{koda}. This fact suggests that AFM and SC have a common background.
In $0 < P < 0.2$ GPa, the marginal AFM is expelled by the onset of the SC below $T_{\rm c}$ at $H$ = 0.
However, when the application of $H$ turns on to suppress the SC, the first-order like transition from the SC to the magnetic {\it A-phase} takes place \cite{Bruls94}.
In $P >$ 0.2 GPa, as a result of the complete suppression of the marginal AFM, the typical HF SC takes place with the line-node gap.

In CeCu$_2$(Si$_{1-x}$Ge$_x$)$_2$, one $4f$ electron per Ce ion plays vital role for both the SC and AFM in $0 \leq x < 0.06$ leading to the novel states of matter. We have proposed that the uniform mixed phase of AFM and SC in the slightly Ge substituted compounds, and the magnetic-field induced  {\it A-phase} for the homogeneous CeCu$_2$Si$_2$ in $0 < P < 0.2$ GPa are accounted for on the basis of an SO(5) theory \cite{kitaoka01,zhang97}.
It is considered that the marginal AFM in $0 \leq x < 0.06$  below $T_{\rm c}$ may be identified as a collective magnetic mode in the uniform mixed phase of AFM and SC and that it turns out to be competitive with the onset of SC in $0 < P < 0.2$ GPa at $H$ = 0. The latter is relevant with the $H$-induced first-order transition from the SC to the {\it A-phase}. Concerning the interplay between AFM and SC, we would propose that the marginal AFM in the SC at $x$ = 0 may correspond to a pseudo Goldstone mode due to the broken U(1) symmetry. Due to the closeness to the magnetic QCP, however, such the gapped mode in the SC should be characterized by an extremely tiny excitation (resonance) energy.

The intimate interplay between SC and AFM found in uniform
CeCu$_2$Si$_2$ has been a long-standing problem - unresolved for over a decade. We have proposed that the SO(5) theory constructed on the basis of quantum-field theory may give a coherent interpretation for these exotic phases found in the HF superconductor CeCu$_2$Si$_2$ \cite{kitaoka01}. In this context, we would suggest that the SC in CeCu$_2$Si$_2$ could be mediated by the same magnetic interaction as leads to the AFM in CeCu$_2$(Si$_{1-x}$Ge$_x$)$_2$. This is in marked contrast to the BCS superconductors in which the pair binding is mediated by phonons $-$ vibrations of the lattice density.

\section{Uniform mixed phase of AFM and SC in CeRhIn$_5$ under $P$}

\subsection{The temperature versus pressure phase diagram}

A new antiferromagnetic HF compound CeRhIn$_5$ undergoes the helical magnetic order at a N\'eel temperature $T_N=3.8$ K with an incommensurate wave vector ${\rm q_M=(1/2, 1/2, 0.297)}$ \cite{Curro}. A neutron experiment revealed the reduced Ce magnetic moments $M_s\sim$ 0.8$\mu_{\rm B}$ \cite{christianson,Bao}. The $P$-induced transition from AFM to SC takes place at a relatively lower critical pressure $P_c=1.63$ GPa and higher $T_c = 2.2$ K than in previous examples \cite{Jaccard92,Mathur98,Grosche96,Movshovic96,Hegger00,Walker97}.
Figure 6 indicates the $P$ vs $T$ phase diagram of CeRhIn$_5$ for AFM and SC that was determined by the In-NQR measurements under $P$.  

\begin{figure}[htbp]
\centering
\includegraphics[width=7.7cm]{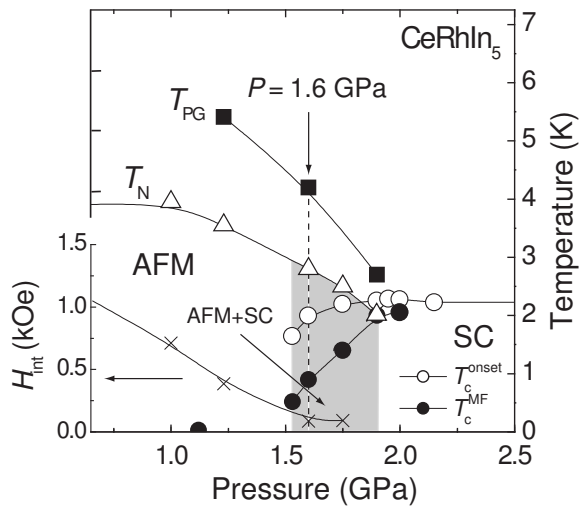}
\caption[]{The $P$ vs $T$ phase diagram for CeRhIn$_5$.
The respective marks denoted by solid square, open triangle and cross correspond to the pseudogap temperature $T_{PG}^{NQR}$, the antiferromagnetic ordering temperature $T_N$ and the internal field $H_{int}$ at the In site. The open and solid circles correspond to the onset temperature $T_c^{onset}$ and $T_c^{MF}$ of superconducting transition (see text). Dotted line denotes the position for $P=1.6$ GPa. Shaded region indicates the coexistent $P$ region of AFM and SC.}
\end{figure}

The NQR study showed that $T_N$ gradually increases up to 4 K as $P$ increases up to $P = 1.0$ GPa and decreases with further increasing $P$ \cite{Mito01,Mito03,Shinji}. In addition, the $T$ dependence of $1/T_1$ probed the pseudogap behavior at $P = 1.23$ and 1.6 GPa \cite{Shinji}. This suggests that CeRhIn$_5$ may resemble other strongly correlated electron systems \cite{Timusk,Kanoda}. Note that the value of bulk superconducting transition temperature $T_c^{MF}$ is progressively reduced as shown by closed circle in figure 6. Apart from the AFM at $P = 2.1$ GPa exceeding $P_c$, {\it $1/T_1$ decreases obeying a $T^3$ law} without the coherence peak just below $T_c$. This indicates that the SC of CeRhIn$_5$ is unconventional with the line-node gap \cite{Mito01,KohoriEPJ00}.  

\begin{figure}[htbp]
\centering
\includegraphics[width=6.5cm]{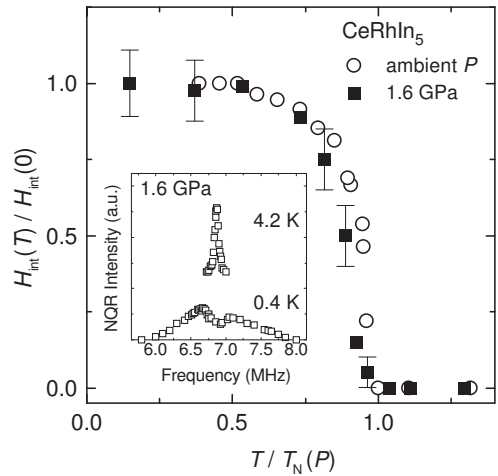}
\caption[]{ Plots of  $H_{int}(T)/H_{int}(0)$ vs $T/T_N$ at $P$ = 0 and 1.6 GPa (see text). Inset shows the $^{115}$In-NQR spectra of 1$\nu_Q$ at $P$ = 1.6 GPa above and below $T_N$ = 2.8 K. }
\end{figure}

\subsection{Gapless magnetic and quasi-particles excitations due to the uniform mixed phase of AFM and SC}

We present microscopic evidence for the exotic SC at the uniform mixed phase of AFM and SC in CeRhIn$_5$ at 
$P = 1.6$ GPa. The inset of figure 7 displays the NQR spectra above and below $T_N$ at $P = 1.6$ GPa. Below $T_N = 2.8$ K, the NQR spectrum splits into two peaks due to the appearance of $H_{int}$ at the In site. This is clear evidence for the occurrence of AFM at $P = 1.6$ GPa. The plots of $H_{int}(T)/H_{int}(0)=M_s(T)/M_s(0)$ vs $(T/T_N)$ at $P = 0$ and 1.6 GPa are compared in figure 7, showing nearly the same behavior. Here $H_{int}(0)$ is the value extrapolated to zero at $T = 0$ K and $M_s(T)$ is the $T$ dependence of spontaneous staggered magnetic moment. The character of AFM at $P = 1.6$ GPa is expected to be not so much different from that at $P = 0$. 

Figure 8 indicates the $T$ dependence of $1/T_1$ at $P = 1.6$ GPa. A clear peak in $1/T_1$ is due to  antiferromagnetic critical fluctuations at $T_N = 2.8$ K. Below $T_N = 2.8$ K, $1/T_1$ continues to decrease moderately down to $T_c^{MF} = 0.9$ K even though passing across $T_c^{onset}\sim 2$ K. This relaxation behavior suggests that SC does not develop following the mean-field approximation below $T_c^{onset}$.  Markedly, $1/T_1$ decreases below $T_c^{MF}$, exhibiting a faint $T^3$ behavior in a narrow $T$ range. With further decreasing $T$, $1/T_1$ becomes proportional to the temperature, indicative of the gapless nature in low-lying excitation spectrum in the coexistent state of SC and AFM on a microscopic level. Thus the $T_1$ measurement has revealed that the intimate interplay between AFM and SC  gives rise to an {\it amplitude fluctuation of superconducting order parameter} between $T_c^{onset}$ and $T_c^{MF}$. 
Such fluctuations may be responsible for the broad transition in resistance and ac-susceptibility ($\chi_{ac}$) measurements. Furthermore, the $T_1T=$ const behavior well below $T_c^{MF}$ evidences the gapless nature in low-lying excitations at the uniform mixed phase of AFM and SC. This result is consistent with those in CeCu$_2$Si$_2$ at the border to AFM \cite{kawasaki01} and a series of CeCu$_2$(Si$_{1-x}$Ge$_2$)$_2$ compounds that show the uniform mixed phase of AFM and SC \cite{kitaoka01,kawasaki02}.  The specific-heat result under $P$, that probed a finite value of its $T$-linear contribution, $\gamma_0$ $\sim$ 100 mJ/molK$^2$ at $P = 1.65$ GPa is now understood due not to a first-order like transition from AFM to SC \cite{Fisher}, but to the gapless nature in the uniform mixed phase of AFM and SC.

\begin{figure}[htbp]
\centering
\includegraphics[width=7cm]{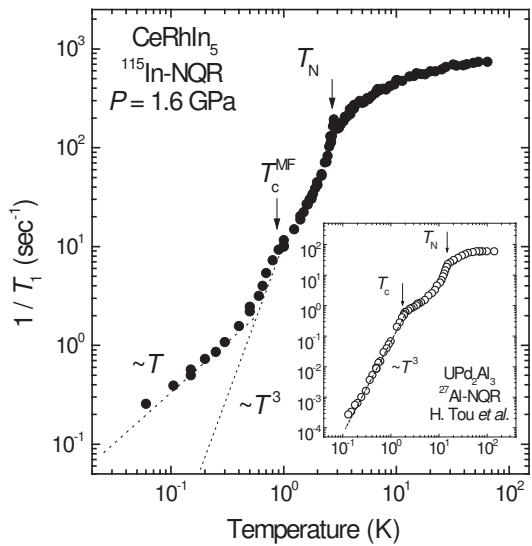}
\caption[]{The $T$ dependence of $1/T_1$ at $P$ = 1.6 GPa.  Both dotted lines correspond to $1/T_1\propto T$ and $1/T_1\propto T^3$.  Inset indicates the $T$ dependence of $^{27}$Al-NQR $1/T_1$ of UPd$_2$Al$_3$ cited  from the literature \cite{Tou}. Dotted line corresponds to $1/T_1\propto T^3$.}
\end{figure}

It is noteworthy that such $T_1T$ = const. behavior is not observed  below $T_c$ at $P$ = 2.1 GPa \cite{Mito01}, consistent with the specific-heat result under $P$ as well \cite{Fisher}. This means that the origin for the $T_1T=$ const. behavior below $T_c^{MF}$ at $P = 1.6$ GPa is not associated with some impurity effect. If it were the case, the residual density of states below $T_c$ should not depend on $P$. 
This novel feature differs from the uranium(U)-based HF antiferromagnetic superconductor UPd$_2$Al$_3$ which has multiple 5$f$ electrons. In UPd$_2$Al$_3$, a superconducting transition occurs at $T_c$ = 1.8 K well below $T_N$ = 14.3 K \cite{Geibel,Steglich2}. As indicated in the inset of figure 8 \cite{Tou}, in UPd$_2$Al$_3$, {\it $1/T_1$ decreases obeying a $T^3$ law over three orders of magnitude} below the onset of $T_c$ without any trace for the $T_1T$ = const. behavior. This is consistent with the line-node gap even in the uniform mixed phase of AFM and SC. 

\subsection{Superconducting fluctuation due to the uniform mixed phase of AFM and SC} 

In order to highlight the novel SC on a microscopic level, the $T$ dependence of $1/T_1T$ is shown in figure 9(a) at $P = 1.6$ GPa in $T = 0.05 - 6$ K and is compared with the $T$ dependence of the resistance $R(T)$ at $P = 1.63$ GPa referred from the literature \cite{Hegger00}. Although each value of $P$ is not exactly the same, they only differ by 2\%.  We remark that the $T$ dependence of $1/T_1T$ points to the pseudogap behavior around $T_{PG}^{NQR}$ = 4.2 K, the AFM at $T_N$ = 2.8 K, and the SC at $T_c^{MF}$ = 0.9 K at which $d\chi_{ac}/dT$ has a peak as seen in figure 9(b). This result itself evidences the uniform mixed state of AFM and SC.  A comparison of $1/T_1T$ with  the $R(T)$ at $P = 1.63$ GPa in figure 9(b) is informative in shedding light on the uniqueness of AFM and SC. Below $T_{PG}^{NQR}$, $R(T)$ starts to decrease more rapidly than a $T$-linear variation extrapolated from a high $T$ side. It continues to decrease across $T_N$ = 2.8 K, reaching the zero resistance at $T_c^{zero}\sim$ 1.5 K.

The resistive superconducting transition width becomes broader. Unexpectedly, $T_c^{onset}\sim$ 2 K, that is defined as the temperature below which the diamagnetism starts to appear, is higher than $T_c^{zero}\sim$ 1.5 K. Any signature for the onset of SC from the $1/T_1$ measurement is not evident in between $T_c^{onset}$ and $T_c^{MF}$, demonstrating that the mean-field type of gap does not grow up down to $T_c^{MF}\sim$ 0.9 K. {\it The existence of fluctuations due to the interplay of AFM and SC is responsible for the broad transition towards the uniform mixed phase of SC and AFM.} 
\begin{figure}[htbp]
\centering
\includegraphics[width=7.7cm]{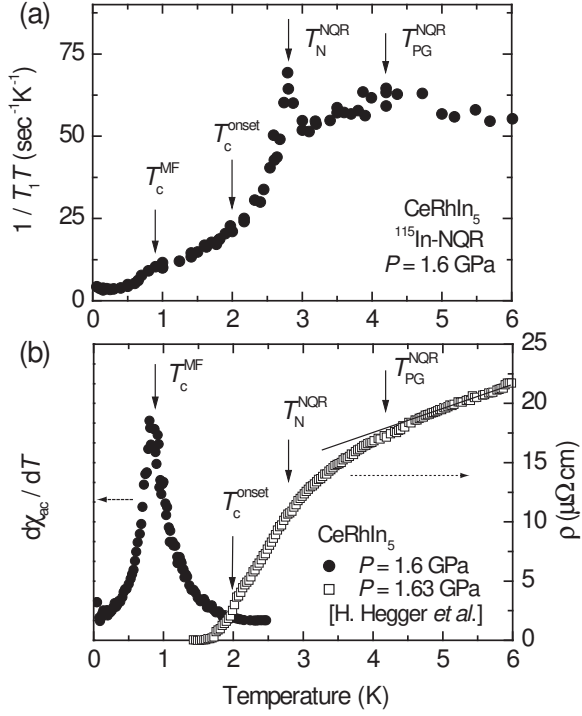}
\caption[]{(a) The $T$ dependence of $1/T_1T$ at $P = 1.6$ GPa. (b) The $T$ dependencies of $d\chi_{ac}/dT$ at $P = 1.6$ GPa and resistance at 
$P = 1.63$ GPa cited  from the literature \cite{Hegger00}. $T_c^{MF}$ and $T_c^{onset}$ correspond to the respective temperatures at which $d\chi_{ac}/dT$ has a peak and below which $\chi_{ac}$ starts to decrease. $T_N$ corresponds to the antiferromagnetic ordering temperature at which $1/T_1T$ exhibits a peak and $T_{PG}^{NQR}$ to the pseudogap temperature below which it starts to decrease. A solid line is an eye guide for the $T$- linear variation in resistance at temperatures higher than $T_{PG}^{NQR}$.}
\end{figure}

\subsection{A novel interplay between AFM and SC}

Recent neutron-diffraction experiment suggests that the size of staggered moment $M_s$ in the AFM is almost independent of $P$ \cite{Bao}. Its relatively large size of moment with $M_s\sim 0.8\mu_B$ seems to support such a picture that the {\it same $f$-electron} exhibits simultaneously itinerant and localized dual nature, because there is only one $4f$-electron per Ce$^{3+}$ ion. In this context, it is natural to consider that the superconducting nature in the uniform mixed phase of AFM and SC belongs to a novel class of phase which differs from the {\it unconventional $d$-wave SC with the line-node gap}. As a matter of fact, a theoretical model has been recently put forth to address the underlying issue in the uniform mixed phase of AFM and SC \cite{Fuseya}.

\section{Emergent phases of SC and AFM on the magnetic criticality in CeIn$_3$}
 
\subsection{The temperature versus pressure phase diagram}

Figure 10 indicates the $P$ vs $T$ phase diagram in CeIn$_3$ around $P_c$. This work has deepened the understanding of the physical properties on the verge of AFM in CeIn$_3$ that exhibits the archetypal phase diagram shown in figure 1(a) \cite{Shinji04}.

The localized magnetic character is robust up to  $P = 1.9$ GPa. The characteristic temperature $T^{*}$, below which the system crosses over to an itinerant magnetic regime, increases dramatically with further increase of $P$.  As a result, the measurements of $1/T_1$ and $\chi_{ac}$ at $P = 2.65$ GPa down to $T = 50$ mK provided the first evidence of unconventional SC at $T_c = 95$ mK in CeIn$_3$, which arises in the HF state fully established below $T_{FL} = 5$ K \cite{Shinji02In3}.

By contrast, the phase separation into AFM and paramagnetism (PM) is evidenced in CeIn$_3$ from the observation of two kinds of NQR spectra in $P = 2.28 - 2.50$ GPa. Nevertheless, it is highlighted that the SC in CeIn$_3$ occurs in both the AFM and PM at $P = 2.43$ GPa. The maximum value of $T_c^{max}=230$ mK is observed for the SC in PM. Markedly, the SC coexisting with AFM emerges below $T_c = 190$ mK. The present results indicate the occurrence of the first-order phase transition from the uniform mixed phase of SC and AFM to the single phase of SC under the HF state of PM around $P_c$. Therefore, a QCP is absent in CeIn$_3$\cite{Shinji04}.

\begin{figure}[h]
\centering
\includegraphics[width=7.7cm]{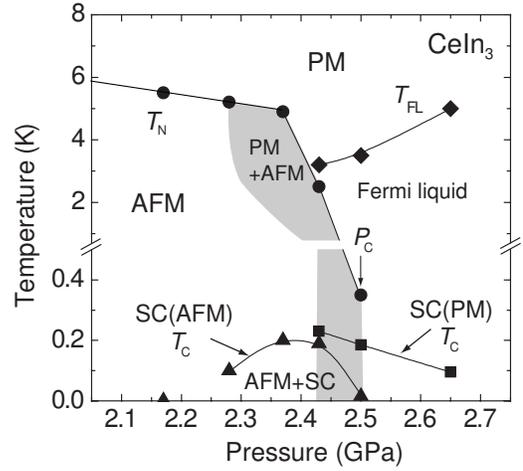}
\caption[]{The $P$ vs $T$ phase diagram of CeIn$_3$ determined from the present experiment. The $P$ and $T$ ranges where the phase separation of AFM and PM occurs are shaded in the figure.}
\end{figure}

CeIn$_3$ forms in the cubic AuCu$_3$ structure and orders antiferromagnetically below $T_N = 10.2$ K at $P = 0$ with an ordering vector {\bf Q} = (1/2,1/2,1/2) and Ce magnetic moment $M_S\sim 0.5\mu_B$, which were determined by NQR measurements\cite{Kohori99,Kohori00} and the neutron-diffraction experiment on single crystals\cite{Knafo}, respectively. The resistivity measurements of CeIn$_3$ have clarified the $P - T$ phase diagram of AFM and SC: $T_N$ decreases with increasing $P$. On the verge of AFM, the SC emerges in a narrow $P$ range of about 0.5 GPa, exhibiting a maximum value of $T_c\sim 0.2$ K at $P_c = 2.5$ GPa where AFM disappears \cite{Walker97,Mathur98,Grosche01,Muramatsu01,Knebel02}.

\begin{figure}[h]
\centering
\includegraphics[width=7.5cm]{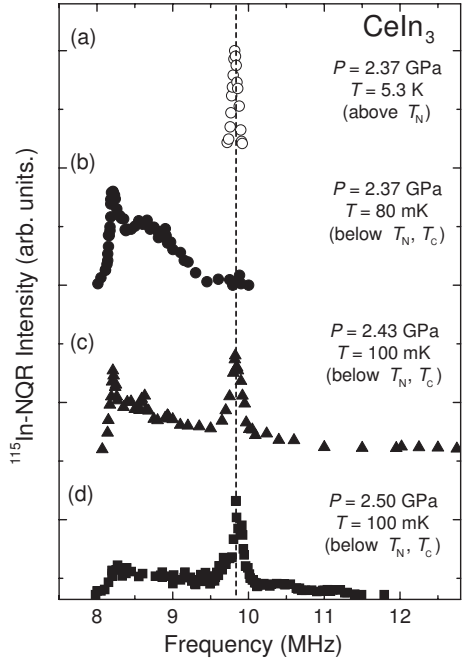}
\caption[]{The $P$ dependence of $^{115}$In NQR spectrum for CeIn$_3$ at $P=2.37$ GPa above $T_N$ (a) and $P$ = 2.37 GPa (b), 2.43 GPa (c) and 2.50 GPa (d) at temperatures lower than the $T_N$ and $T_c$. The dotted line indicates the peak position at which the NQR spectrum is observed for PM.}
\end{figure}

\subsection{Evidence of the phase separation into AFM and PM near their boundary}

Figure 11 shows the NQR spectra of 1$\nu_{Q}$ transition for the PM at (a) $P = 2.37$ GPa and for temperatures lower than $T_N$ and $T_c$ at (b) $P = 2.37$ GPa, (c) $P = 2.43$ GPa and (d) $P = 2.50$ GPa. Note that the 1$\nu_Q$ transition can sensitively probe the appearance of internal field $H_{int}(T)$ associated with even tiny Ce ordered moments on the verge of AFM. As a matter of fact, as seen in figures 11(a) and (b), a drastic change in the NQR spectral shape is observed due to the occurrence of $H_{int}(T)$ at the In nuclei below $T_N$. By contrast, the spectra at (c) $P = 2.43$ GPa and  (d) $P = 2.50$ GPa include two kinds of spectra arising from AFM and PM provides firm microscopic evidence for the emergence of magnetic phase separation. The volume fraction of AFM at $T = 100$ mK at each pressure is plotted as function of $P$, as seen in figure 12(a) (solid triangles). It should be noted that, as shown in figure 10, the phase separations at $P = 2.28$ and 2.37 GPa are observed only between $T = 3$ K and $T_N = 5.2$ K and between $T = 1$ K and $T_N = 4.9$ K, respectively.
 
\begin{figure}[htbp]
\centering
\includegraphics[width=7.5cm]{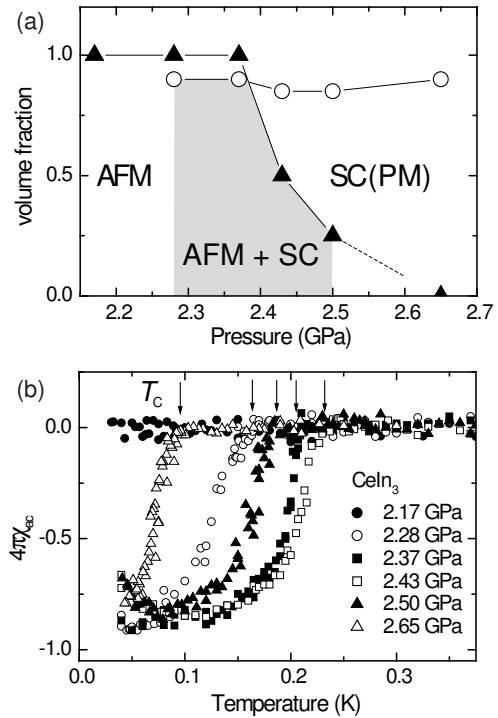}
\caption[]{(a) The $P$ dependence of the volume fraction of AFM(solid triangles) and SC in PM(open circles). The volume fraction of AFM is evaluated from the integration of the NQR spectrum over the frequency at $T$ = 100 mK at each pressure. The superconducting volume fraction at each pressure is estimated by comparing the value of $\chi_{ac}$ at the lowest temperature for CeIn$_3$ with that for the HF superconductor CeIrIn$_5$\cite{Shinji02In3}. The phase separation of AFM and PM takes place at $P = 2.28-2.50$ GPa. Note that the uniform mixed phase of SC and AFM emerges  in the shaded region. (b) The $T$ dependence of $\chi_{ac}$ in $P = 2.17 - 2.65$ GPa. The solid arrow indicates the onset $T_c$ for SC.
}
\end{figure}

Figure 12(b) shows the $T$ dependence of $\chi_{ac}$ for CeIn$_3$ under various values of $P$.  Even though the magnetic phase separation takes place, as shown by shading in figure 10, the clear decrease in $\chi_{ac}$ points to the bulk nature of SC at $P = 2.28 - 2.65$ GPa. It is, however, noteworthy that there is no indication of SC down to $T$ = 30 mK at $P$ = 2.17 GPa where no magnetic phase separation is observed at all against $P$ and $T$. When taking into account the fact that the value of $T_c$ reaches a maximum at $P= 2.43$ GPa where the volume fraction of AFM and PM remains comparable, it is expected that the uniform mixed phase of SC and AFM takes place only in the $P$ region where the phases of AFM and PM are separated. The results of the $P$ dependence of NQR spectrum and $\chi_{\rm ac}$ strongly suggested that the uniform mixed phase of AFM and SC is separated from the phase of HF SC under PM in the shaded region in figure 12(a). The present results revealed that there emerges the first-order transition from the uniform mixed phase of SC and AFM to the phase of HF SC under PM. Note that the SC takes place under both the backgrounds of AFM and PM, although the superconducting characteristics are significantly different. 

\subsection{The novel phases of SC under both the backgrounds of AFM and PM}

The uniform mixed phase of AFM and SC is corroborated by direct evidence from the $T$ dependence of $1/T_1T$ that can probe the low-lying excitations due to quasiparticles in SC and the magnetic excitations in AFM.  Figure 13 shows the drastic evolution in the $P$ and $T$ dependencies of $1/T_1T$ for AFM (solid symbols) and PM (open symbols) at $P$ = 2.17, 2.28, 2.43 and 2.50 GPa. Here, $T_N$ was determined as the temperature below which the NQR intensity for PM decreases due to the emergence of AFM associated with the magnetic phase separation. The $T_1$ for AFM and PM is separately measured at respective NQR peaks which are clearly distinguished from each other, as shown in figures 11(b), 11(c) and 11(d). Thus, the respective $T_c^{AFM}$ and $T_c^{PM}$ for AFM and PM are determined as the temperature below which $1/T_1T$ decreases markedly due to the opening of superconducting gap. These results verify that the uniform mixed phase of AFM and SC takes place on a microscopic level in $P = 2.28 - 2.50$ GPa.

\begin{figure}[htbp]
\centering
\includegraphics[width=8.5cm]{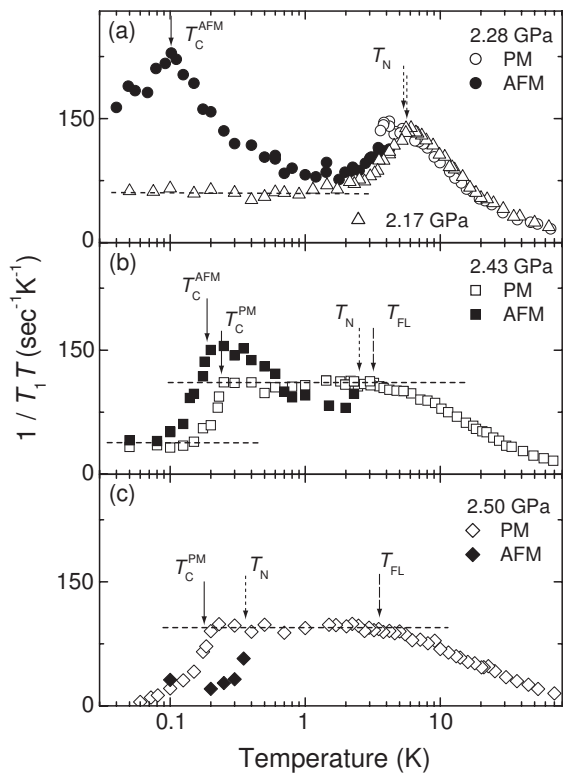}
\caption[]{$T$ dependence of $ ^{115}(1/T_1T)$ in CeIn$_3$ at $P = 2.17$ and 2.28 GPa (a), 2.43 GPa (b) and 2.50 GPa (c). Open and solid symbols indicate the data for PM and AFM measured at $\sim$ 9.8 MHz and $\sim$ 8.2 MHz, respectively. The solid arrow indicates the respective superconducting transition temperatures $T_c^{PM}$ and $T_c^{AFM}$ for PM and AFM.  The dotted and dashed arrows indicate, respectively, the $T_N$ and the characteristic temperature $T_{FL}$ below which the $T_1T = $const. law (dotted line) is valid, which is characteristic of the Fermi-liquid state.}
\end{figure}

In the PM at $P = 2.43$ and 2.50 GPa, the $1/T_1T$ = const. relation is valid below $T_{FL}\sim 3.2$ K and $\sim 3.5$ K, respectively, which indicates that the Fermi-liquid state is realized. This result is in good agreement with that of the previous resistivity measurement from which the $T^2$ dependence in resistance was confirmed \cite{Knebel02}. Note that the $1/T_1$ for the SC in the PM at $P = 2.50$ GPa follows a $T^3$ dependence below $T_c = 190$ mK, consistent with the line-node gap model characteristic for unconventional HF SC.

As shown in figure 13(a), no phase separation occurs below $T_N = 5.5$ K at $P = 2.17$ GPa, and $1/T_1T = $ const. behavior is observed well below $T_N$. It is also observed in the previous NQR measurements at $P = 0$ \cite{Kohori99,Kohori00}. The $T$ dependence of $1/T_1T$ at $P = 2.28$ GPa resembles the behavior at $P = 2.17$ GPa above $T \sim$ 1 K. However, at $P = 2.28$ GPa, the phase separation of AFM and PM occurs in the small $T$ window between $T_N = 5.2$ K and 3 K. In contrast to $1/T_1T = $ const. behavior well below $T_N$ at $P = 2.17$ GPa, the $1/T_1T$ at $P = 2.28$ GPa continues to increase upon cooling below $T_N$ and exceeds the values of around $T_N$ in spite of antiferromagnetic spin polarization being induced. These results suggest that the occurrence of magnetic phase separation is closely responsible for the onset of SC, and yields the large enhancement of the low-lying magnetic excitations for AFM. This feature is also seen for the AFM at $P = 2.43$ GPa where $1/T_1T$ is larger than the value for PM, as shown in figure 13(b). It is still unknown why such low-lying magnetic excitations continue to be enhanced well below $T_N$. Some spin-density fluctuations may be responsible for this feature in association with the phase separation of AFM and PM. In this context, CeIn$_3$ is not in a magnetically soft electron liquid state \cite{Mathur98}, but instead, the relevant magnetic excitations, such as spin-density fluctuations, induced by the first-order transition from AFM to PM might mediate attractive interaction. Whatever its pairing mechanism is at $P = 2.28$ GPa where AFM is realized over the whole sample below $T = 3$ K, the clear decrease in $1/T_1T$ and $\chi_{ac}$ provide convincing evidence for the uniform mixed phase of AFM and SC in CeIn$_3$ at $P = 2.28$ GPa.

Further evidence for the new type of SC uniformly mixed with AFM was obtained from the results at $P = 2.43$ GPa, as indicated in figure 13(b).  At temperatures lower than the respective values of $T_c^{PM}$ = 230 mK and $T_c^{AFM}$ = 190 mK for PM and AFM, unexpectedly, the magnitude of $1/T_1T = $ const. coincide with one another for both the phases that are magnetically separated into AFM and PM with the respective different values of $T_c$. This means that  the quasi-particle excitations for the uniform mixed phase of SC and AFM may be the same in origin as for the phase of SC in PM. How does this happen? It may be possible that both the phases are in a dynamically separated regime with time scales smaller than the inverse of NQR frequency so as to make each superconducting phase for the SC under AFM and for the PM uniform. In this context, the observed magnetically separated  phases and the relevant phases of SC may belong to new phases of matter.

We have provided evidence for the phase separation of AFM and PM (shaded area in figure 10) and the new type of SC uniformly mixed with the AFM near $P_c$ in CeIn$_3$. It has been found that the highest value of $T_c = 230$ mK in CeIn$_3$ is observed for the PM at $P = 2.43$ GPa where the volume fraction of AFM and PM becomes almost the same. The present experiments have revealed that this new type of SC is mediated by a novel pairing interaction associated with the magnetic phase separation. We propose that {\it the magnetic excitations, such as spin-density fluctuations, induced by the first-order magnetic phase transition might mediate attractive interaction to form Cooper pairs in CeIn$_3$; this is indeed a new type of pairing mechanism}.

\section{Towards understanding of universal concept for the superconductivity in heavy-fermions systems}

The SC in HF compounds has not yet been explained from the microscopic point of view, mainly due to the strong correlation effect and the complicated band structures. An essential task seems to identify the residual interaction between quasi-particles through analyzing the effective $f$-band model by choosing dominant bands \cite{Yanase}. Here, we have demonstrated that HF superconductors possess a great variety of ground states at the boundary between SC and AFM with anomalous magnetic and superconducting properties.

A genuine uniform mixed phase of AFM and SC has been observed in CeCu$_2$Si$_2$ and CeRhIn$_5$ in the pressure ($P$) versus temperature ($T$) phase diagram through the extensive and precise NQR measurements under $P$. In other strongly correlated electron systems, the SC appears near the boundary to the AFM. Even though the underlying solid state chemistries are rather different, the resulting phase diagrams are strikingly similar and robust. This similarity suggests that the overall feature of all these phase diagrams is controlled by a single energy scale. In order to gain an insight into the interplay between AFM and SC, here, we try to focus on a particular theory, which unifies the AFM and SC of the heavy-fermion systems based on an SO(5) theory, because symmetry unifies apparently different physical phenomena into a common framework as all fundamental laws of Nature \cite{SO(5)review}. 
The uniform mixed phase diagram of AFM and SC and the exotic SC affected by strong antiferromagnetic fluctuations could be understood in terms of an SO(5) superspin picture \cite{kitaoka01,zhang97}.

By contrast, CeIn$_3$ has revealed the $P$ induced first-order transition from  AFM to PM as functions of pressure and temperature near their boundary.  Unexpectedly, however, the SC is robust under the phase separation into AFM and PM, even though the superconducting characteristics in the AFM strikingly differ from those in the PM. The uniform mixed phase of AFM and SC has been also observed, however, neither superconducting fluctuations nor the development of low-lying magnetic excitations have been revealed. These features differ significantly from those observed in CeCu$_2$Si$_2$ and CeRhIn$_5$. Instead, the spin-density fluctuations develop as temperature goes down far below $T_N$, leading to the onset of  the SC under the background of AFM which is separated from the PM.

These new phenomena observed in CeIn$_3$ should be understood in terms of a quantum phase separation because these new phases of matter are induced by applying pressure. In Fermion systems, if the magnetic critical temperature at the termination point of the first-order transition is suppressed at $P_c$, the diverging magnetic density fluctuations inherent at the critical point from the magnetic to paramagnetic transition become involved in the quantum Fermi degeneracy region. The Fermi degeneracy by itself generates various instabilities called as the Fermi surface effects, one of which is a superconducting transition. On the basis of a general argument on quantum criticality, it is shown that the coexistence of the Fermi degeneracy and the critical density fluctuations yield a new type of quantum criticality \cite{Imada}. In this context, the results on CeIn$_3$ deserve further theoretical investigations. 

We believe that the results presented here on CeCu$_2$Si$_2$, CeRhIn$_5$ and CeIn$_3$ provide vital clue to unravel the essential interplay between the AFM and the SC, and to extend the universality of the understanding on the SC in strongly correlated electron systems.

\section{Acknowledgement}
These works are currently stimulated in collaboration with C. Geibel and F. Steglich on the study of CeCu$_2$Si$_2$ and with Prof. Y. \=Onuki and his many coworkers on the studies of CeRhIn$_5$ and CeIn$_3$. These works were supported by the COE Research grant from MEXT of Japan (Grant No.\ 10CE2004).\\


\end{document}